\documentclass[11pt]{article}
\usepackage{amssymb}
\usepackage{epsfig}

\newcommand{\BE}{\begin{equation}}
\newcommand{\EE}{\end{equation}}
\begin{document}
\begin{titlepage}

\vspace*{1mm}
\begin{center}

   {\LARGE{\bf Indications for a preferred reference frame  \\
from an ether-drift experiment }}

\vspace*{14mm}
{\Large  M. Consoli and E. Costanzo}
\vspace*{4mm}\\
{\large
Istituto Nazionale di Fisica Nucleare, Sezione di Catania \\
Dipartimento di Fisica e Astronomia dell' Universit\`a di Catania \\
Via Santa Sofia 64, 95123 Catania, Italy}
\end{center}
\begin{center}
{\bf Abstract}
\end{center}
We present a fully model-independent analysis of the extensive
observations reported by a recent ether-drift experiment in Berlin.
No a priori assumption is made on the nature of a hypothetical
preferred frame. We find a remarkable consistency with an Earth's
cosmic motion exhibiting an average declination angle $|\gamma|\sim
43^o$ and with values of the RMS anisotropy parameter
$(1/2-\beta+\delta)$ that are one order of magnitude larger than the
presently quoted ones. This might represent the first modern
indication for a preferred frame and for a non-zero anisotropy of
the speed of light.\vskip 35 pt PACS: 03.30.+p, 01.55.+b \vskip 150
pt Submitted to Physical Review Letters
\end{titlepage}

%\setcounter{equation}{0}
%\section{Introduction}

The present generation of ether-drift experiments, combining the
possibility of active rotations of the apparatus with the use of
cryogenic optical resonators, is currently pushing the relative
accuracy of the measured frequency shifts to the level ${\cal
O}(10^{-16})$. Therefore, it becomes important that such a high
precision is not obscured by model-dependent assumptions in the
analysis of the data that might introduce uncontrolled errors in the
determination of the physical parameters. In this Letter we'll
present a fully model-independent analysis of the extensive
observations reported in Ref.\cite{peters} without constraining a
hypothetical preferred frame to coincide with the CMB. By removing
this assumption, the data provide consistent indications for the
existence of a different type of preferred frame and for an
anisotropy of the two-way speed of light that is one order of
magnitude larger than the presently quoted one.

The starting point for our analysis is the expression for the
relative frequency shift of two orthogonal optical resonators at a
given time $t$. This is expressed as
\BE
\label{basic2}
      {{\delta \nu [\theta(t)]}\over{\nu_0}} =
      {S}(t)\sin 2\theta(t) +
      {C}(t)\cos 2\theta(t)
\EE
where $\theta(t)$ is the angle of rotation of the apparatus. The Fourier
expansion of the two
amplitudes $S(t)$ and $C(t)$ is predicted to be
\BE
\label{amorse1}
      {S}(t) =
      {S}_{s1}\sin\tau +{S}_{c1} \cos\tau
       + {S}_{s2}\sin(2\tau) +{S}_{c2} \cos(2\tau)
\EE
\BE
\label{amorse2}
      {C}(t) = {C}_0 +
      {C}_{s1}\sin\tau +{C}_{c1} \cos\tau
       + {C}_{s2}\sin(2\tau) +{C}_{c2} \cos(2\tau)
\EE
where $\tau=\omega_{\rm sid}t$ is the sidereal time of the observation
in degrees and $\omega_{\rm sid}\sim {{2\pi}\over{23^{h}56'}}$.

Introducing the colatitude of the laboratory $\chi \sim 37.5^o$, and
the unknown average velocity, right ascension and declination of the
cosmic motion with respect to a hypothetical preferred frame,
(respectively $v$, $\alpha$ and $\gamma$) one finds the expressions
reported in Table I of Ref.~\cite{peters},
\BE
\label{C0}
      {C}_0 =- (1/2-\beta+\delta) {{ v^2 }\over{c^2}}
      {{\sin^2\chi}\over{8}} (3 \cos 2{\gamma} -1) ,
\EE
\BE
\label{CS1}
      {C}_{s1}= {{1}\over{4}}(1/2-\beta+\delta) {{ v^2 }\over{c^2}}
      \sin 2{\gamma} \sin{\alpha} \sin 2\chi ,
\EE
\BE
\label{CC1}
      {C}_{c1}={{1}\over{4}}(1/2-\beta+\delta)
      {{ v^2 }\over{c^2}} \sin 2{\gamma}
      \cos{\alpha} \sin 2\chi ,
\EE
\BE
\label{CS2}
      {C}_{s2} = {{1}\over{4}}(1/2-\beta+\delta)
      {{ v^2 }\over{c^2}} \cos^2{\gamma}
      \sin2{\alpha}  (1+ \cos^2\chi) ,
\EE
\BE
\label{CC2}
      {C}_{c2} = {{1}\over{4}}(1/2-\beta+\delta)
      {{ v^2 }\over{c^2}} \cos^2{\gamma}
      \cos2{\alpha} (1+ \cos^2\chi)
\EE
\vfill \eject
 where $(1/2-\beta+\delta)$ indicates the
Robertson-Mansouri-Sexl \cite{robertson,mansouri} (RMS) anisotropy
parameter. The corresponding $S-$quantities are also given by
${S}_{s1}=-{C}_{c1}/\cos\chi$, ${S}_{c1}={C}_{s1}/\cos\chi$,
${S}_{s2}= -{{2\cos\chi}\over{1+\cos^2\chi}}{C}_{c2}$ and ${S}_{c2}=
{{2\cos\chi}\over{1+\cos^2\chi}}{C}_{s2}$.

The experimental data reported in Ref.\cite{peters} refer to 15
short-period observations performed
from December 2004 to April 2005. As suggested by the same authors,
it is safer to concentrate on
the observed time modulation of the signal, i.e. on the
quantities ${C}_{s1},{C}_{c1},{C}_{s2},{C}_{c2}$ (and on
their ${S}$-counterparts). In fact, the constant components
$ \bar{C}={C}_0 $ and $\bar { S}\equiv S_0$
are most likely affected by spurious systematic effects such as thermal
drift (see also the discussion in Ref.~\cite{schiller}).

Since the individual determinations of the various parameters for each of the
15 short-period observations were not explicitely given by the authors of
Ref.\cite{peters}, we have
extracted these values from their Fig.3. The relevant numbers are
reported in our Table 1 and Table 2.
\begin{table*}
\caption{The experimental data
for the $C-$coefficients
as extracted from Fig.3 of Ref.[1]. }
\begin{center}
\begin{tabular}{clll}
\hline\hline
$C_{s1}[{\rm x}10^{-16}]$ &
$C_{c1}[{\rm x}10^{-16}]$ &
$C_{s2}[{\rm x}10^{-16}]$ &
$C_{c2}[{\rm x}10^{-16}]$   \\
\hline
$-2.7\pm4.5$ &  $5.3\pm 4.8$ &  $-3.2\pm4.7$ &  $1.2\pm 4.2$ \\
$-18.6\pm6.5$ &  $8.9\pm 6.4$ &  $-11.4\pm6.5$ &  $-5.0\pm 6.4$ \\
$-0.7\pm3.9$ &  $5.3\pm 3.6$ &  $5.0\pm3.5$ &  $1.6\pm 3.8$ \\
$6.1\pm4.6$ &  $0.0\pm 4.8$ &  $-8.1\pm4.8$ &  $-4.0\pm 4.6$ \\
$2.0\pm8.6$ &  $1.3\pm 7.7$ &  $16.1\pm8.0$ &  $-3.3\pm 7.2$ \\
$3.0\pm5.8$ &  $4.6\pm 5.9$ &  $8.6\pm5.9$ &  $-6.9\pm 5.9$ \\
$0.0\pm5.4$ &  $-9.5\pm 5.7$ &  $-5.5\pm5.6$ &  $-3.5\pm 5.4$ \\
$-1.1\pm8.1$ &  $11.0\pm 7.9$ &  $0.9\pm8.3$ &  $18.6\pm 7.9$ \\
$8.6\pm6.5$ &  $2.7\pm 6.7$ &  $4.3\pm6.5$ &  $-12.4\pm 6.4$ \\
$-4.8\pm4.8$ &  $-5.1\pm 4.8$ &  $3.8\pm4.7$ &  $-5.2\pm 4.7$ \\
$5.7\pm3.2$ &  $3.0\pm 3.4$ &  $-6.3\pm3.2$ &  $0.0\pm 3.5$ \\
$4.8\pm8.0$ &  $0.0\pm 7.0$ &  $0.0\pm7.6$ &  $1.5\pm 7.7$ \\
$3.0\pm4.3$ &  $-5.9\pm 4.3$ &  $-2.1\pm4.4$ &  $14.1\pm 4.3$ \\
$-4.5\pm4.4$ &  $-2.3\pm 4.5$ &  $4.1\pm4.3$ &  $3.2\pm 4.3$ \\
$0.0\pm3.6$ &  $4.6\pm 3.4$ &  $0.6\pm3.2$ &  $4.9\pm 3.3$ \\
\hline\hline
\end{tabular}
\end{center}
\end{table*}
\vfill \eject

\begin{table*}
\caption{The experimental data for the $S-$coefficients as extracted
from Fig. 3 of Ref.[1].}
\begin{center}
\begin{tabular}{clll}
\hline\hline $S_{s1}[{\rm x}10^{-16}]$ & $S_{c1}[{\rm x}10^{-16}]$ &
$S_{s2}[{\rm x}10^{-16}]$ &
$S_{c2}[{\rm x}10^{-16}]$   \\
\hline
$11.2\pm4.7$ &  $11.9\pm 4.9$ &  $1.8\pm4.9$ &  $0.8\pm 4.5$ \\
$1.8\pm6.5$ &  $-4.3\pm 6.5$ &  $6.4\pm6.4$ &  $1.8\pm 6.4$ \\
$-3.3\pm3.8$ &  $2.9\pm 3.8$ &  $-5.9\pm3.8$ &  $4.6\pm 4.0$ \\
$12.7\pm5.1$ &  $14.3\pm 5.5$ &  $-1.9\pm5.3$ &  $-3.3\pm 5.1$ \\
$4.7\pm8.4$ &  $-6.9\pm 7.3$ &  $-1.8\pm8.0$ &  $-7.8\pm 7.0$ \\
$5.2\pm5.8$ &  $-3.0\pm 5.9$ &  $7.1\pm5.9$ &  $-5.9\pm 5.8$ \\
$11.1\pm5.3$ &  $-13.4\pm 5.4$ &  $-4.5\pm5.5$ &  $-9.8\pm 5.5$ \\
$-12.1\pm8.9$ &  $0.0\pm 8.8$ &  $-3.1\pm9.0$ &  $1.4\pm 8.9$ \\
$-4.8\pm6.3$ &  $6.5\pm 6.4$ &  $-8.1\pm6.3$ &  $3.5\pm 6.5$ \\
$9.8\pm5.0$ &  $4.8\pm 5.0$ &  $1.9\pm5.0$ &  $-9.2\pm 4.8$ \\
$0.0\pm3.2$ &  $-3.9\pm 3.6$ &  $1.0\pm3.1$ &  $-2.2\pm 3.4$ \\
$-12.7\pm7.7$ &  $8.5\pm 6.8$ &  $-8.3\pm7.2$ &  $-7.1\pm 7.4$ \\
$-7.9\pm4.7$ &  $-4.3\pm 4.8$ &  $-1.9\pm4.8$ &  $-6.2\pm 4.7$ \\
$16.1\pm4.9$ &  $12.0\pm 5.2$ &  $2.9\pm4.9$ &  $-9.6\pm 4.8$ \\
$13.9\pm3.9$ &  $-7.0\pm 3.4$ &  $-3.3\pm3.5$ &  $3.0\pm 3.6$ \\
\hline\hline
\end{tabular}
\end{center}
\end{table*}

For our analysis, rather than using the individual $C$ and $S$
coefficients themselves, we shall work with the combinations
\BE
\label{csid}
      {C}_{11}\equiv \sqrt{{C}^2_{s1}
      + {C}^2_{c1}}
\EE
\BE
\label{c2sid}
      {C}_{22}\equiv \sqrt{{C}^2_{s2}
      + {C}^2_{c2}}
\EE
 \BE \label{ssid}
      {S}_{11}\equiv \sqrt{{S}^2_{s1}
      + {S}^2_{c1}}
\EE
\BE
\label{s2sid}
      {S}_{22}\equiv \sqrt{{S}^2_{s2}
      + {S}^2_{c2}}
\EE
\vfill \eject
This is useful to reduce the model dependence in the
analysis of the data. In this way, in fact, the right ascension
${\alpha}$ drops out from the theoretical predictions that will only
depend on $|{\gamma}|$, and the overall normalization factor \BE
\label{cappa}
 K=|(1/2-\beta+\delta)|{{v^2}\over{c^2}}
\EE The experimental values for these auxiliary quantities are shown
in our Table 3 (for simplicity we report symmetrical errors).

\begin{table*}
\caption{ The experimental values for the combinations of $C-$ and
$S-$ coefficients defined in Eqs.(\ref{csid})-(\ref{s2sid}) as
obtained from our Table 1 and Table 2. }.
\begin{center}
\begin{tabular}{clll}
\hline\hline $ {C}_{11}  [{\rm x}10^{-16}]$ & $ {C}_{22}  [{\rm
x}10^{-16}]$ & $ {S}_{11}  [{\rm x}10^{-16}]$ &
$ {S}_{22}  [{\rm x}10^{-16}]$ \\
\hline
$5.9\pm4.7$ &  $3.5\pm 4.6$ &  $16.3\pm4.8$ &  $2.0\pm 4.9$ \\
$20.6\pm6.4$ &  $12.5\pm 6.5$ &  $4.6\pm6.5$ &  $6.6\pm 6.4$ \\
$5.3\pm3.6$ &  $5.3\pm 3.6$ &  $4.4\pm3.8$ &  $7.5\pm 3.8$ \\
$6.1\pm4.6$ &  $9.0\pm 4.8$ &  $19.1\pm5.3$ &  $3.8\pm 5.1$ \\
$2.4\pm8.4$ &  $16.5\pm 8.0$ &  $8.4\pm7.7$ &  $8.0\pm 7.1$ \\
$5.5\pm5.9$ &  $11.0\pm 5.9$ &  $6.0\pm5.9$ &  $9.2\pm 5.9$ \\
$9.5\pm5.7$ &  $6.5\pm 5.5$ &  $17.4\pm5.4$ &  $10.7\pm 5.5$ \\
$11.0\pm7.9$ &  $18.7\pm 7.9$ &  $12.1\pm8.9$ &  $3.4\pm 9.0$ \\
$9.1\pm6.5$ &  $13.1\pm 6.4$ &  $8.1\pm6.4$ &  $8.8\pm 6.4$ \\
$7.0\pm4.8$ &  $6.5\pm 4.7$ &  $10.9\pm5.0$ &  $9.4\pm 4.8$ \\
$6.4\pm3.1$ &  $6.3\pm 3.2$ &  $3.9\pm3.6$ &  $2.4\pm 3.4$ \\
$4.8\pm8.0$ &  $1.5\pm 7.7$ &  $15.3\pm7.4$ &  $10.9\pm 7.3$ \\
$6.6\pm4.3$ &  $14.3\pm 4.3$ &  $9.0\pm4.7$ &  $6.5\pm 4.7$ \\
$5.1\pm4.5$ &  $5.2\pm 4.3$ &  $20.0\pm5.0$ &  $10.0\pm 4.8$ \\
$4.6\pm3.4$ &  $5.0\pm 3.3$ &  $15.6\pm3.8$ &  $4.4\pm 3.5$ \\
\hline\hline
\end{tabular}
\end{center}
\end{table*}
\vfill \eject Thus, we obtain the relations
\BE
\label{sid1}
      {C}_{11}= {{1}\over{4}} K
      |\sin 2\gamma | \sin 2\chi
\EE
and
\BE
\label{sid2}
      {C}_{22}= {{1}\over{4}} K \cos^2\gamma
      (1+ \cos^2\chi) .
\EE
The corresponding ${S}$-coefficients are also predicted as
${S}_{11}= {C}_{11}/\cos\chi$ and ${S}_{22}=
{{2\cos\chi}\over{1+\cos^2\chi}}$ ${C}_{22}$.

As one can check, the values reported in Table 3 show a good
consistency. Thus we have computed the weighted averages obtaining
the following results \BE \langle{C}_{11} \rangle= (6.7 \pm
1.2)\cdot 10^{-16}~~~~~~ \langle{C}_{22} \rangle= (7.6 \pm 1.2)\cdot
10^{-16} \EE \BE \langle{S}_{11} \rangle= (11.0 \pm 1.3)\cdot
10^{-16}~~~~~~ \langle{S}_{22} \rangle= (6.3 \pm 1.3)\cdot 10^{-16}
\EE Now, the measured ratio ${{\langle S_{11} \rangle }\over{\langle
C_{11}\rangle }}=1.64\pm 0.36$ is consistent with the theoretical
prediction $1/\cos\chi \sim 1.26$ and the measured ratio ${{\langle
S_{22} \rangle }\over{\langle C_{22}\rangle }}=0.83\pm 0.21$ is
consistent with the theoretical prediction
${{2\cos\chi}\over{1+\cos^2\chi}}\sim 0.97$.

We can thus proceed and obtain 4 independent determinations of the
average $|\gamma|$ from the 4 ratios $R_1={{ \langle{C}_{11} \rangle
}\over{ \langle{C}_{22}\rangle }} $, $R_2={{\langle{C}_{11} \rangle
}\over{ \langle{S}_{22} \rangle}}$, $R_3={{\langle{S}_{11} \rangle
}\over{ \langle{S}_{22} \rangle}}$, $R_4={{\langle{S}_{11} \rangle
}\over{ \langle{C}_{22} \rangle}}$. The results are the following
(for simplicity we report symmetrical errors): 1) from $R_1$ one
gets $|\gamma|=36^o\pm 7^o$, 2) from $R_2$ one gets
 $|\gamma|=41^o\pm 8^o$,
3) from $R_3$ one gets $|\gamma|=49^o\pm 7^o$, 4) from $R_4$ one
gets $|\gamma|=44^o\pm 6^o$. As one can see, all data are well
consistent with an average value \BE |\gamma|\sim  43^o \pm 3^o \EE

After having determined the value of the average declination, we can
finally fix $|\gamma|\sim 43^o$ everywhere and obtain 4 estimates of
the overall normalization factor $K$. The results are the following:
1) from $\langle{C}_{11} \rangle$ one gets $K=(28\pm 5)\cdot
10^{-16}$, 2) from $\langle{C}_{22} \rangle$ one gets $K=(35\pm
6)\cdot 10^{-16}$, 3) from $\langle{S}_{11} \rangle$ one gets
$K=(36\pm 4)\cdot 10^{-16}$, 4) from $\langle{S}_{22} \rangle$ one
gets $K=(30\pm 6)\cdot 10^{-16}$. As one can see, all data are well
consistent with an average value \BE \label{kappa2} K\sim (33\pm
3)\cdot 10^{-16}\EE

Notice the huge difference with respect to the analysis of
Ref.\cite{peters} where the Earth's motion with respect to the CMB
($v\sim$370 km/s, $\alpha\sim 168^o$, $\gamma\sim -6^o$) was assumed
from the very beginning in the analysis of the data. In this case,
fixing $v\sim$ 370 km/s and taking the value of the RMS parameter
from Ref.\cite{peters} $|(1/2-\beta+\delta)|\sim (2\pm 2)\cdot
10^{-10}$, one would rather predict the value $K\sim (3\pm 3)\cdot
10^{-16}$ which is one order of magnitude smaller than the value
reported in our Eq.(\ref{kappa2}). Equivalently, assuming
tentatively the range of velocities 230 km/s $\leq v \leq$ 370 km/s,
to account for most cosmic motions of the Solar System (within the
Galaxy, relatively to the centroid of the Local Group, relatively to
the CMB,...) our result Eq.(\ref{kappa2}) implies a RMS parameter
\BE \label{ourbeta} 20 \cdot 10^{-10} \leq |(1/2-\beta+\delta)| \leq
60 \cdot 10^{-10} \EE

To push further the analysis, one might attempt to use the above
results to constrain the average value of the right ascension
$\alpha$  from the individual average values of the $C-$ and $S-$
coefficients reported in Ref.\cite{peters}. However, this is not
possible at the present. In fact, in order to consider the
individual coefficients one should fix the sign of the declination
angle (i.e. $\gamma=\pm 43^o$) and the sign of the RMS anisotropy
parameter. These have to be combined with the 4 possible choices of
$\alpha$ ($\sin \alpha = \pm |\sin \alpha|$ and $\cos \alpha=\pm
|\cos\alpha|$). The 16 different alternatives cannot be fully
exploited since, at the present, among the 8 average values of the
$C-$ and $S-$ coefficients reported in Ref.\cite{peters}, only
$\langle S_{s1}\rangle $ and $\langle S_{c2}\rangle $ are non-zero
beyond the one-sigma level.

Summarizing: we have presented a fully model-independent analysis of
the extensive ether-drift observations reported in
Ref.\cite{peters}. Without constraining a hypothetical preferred
frame to coincide with the CMB, the experimental data select a
definite (absolute) value of the average declination angle
$|\gamma|\sim 43^o$ and a RMS parameter lying in the typical range
$20 \cdot 10^{-10} \leq |(1/2-\beta+\delta)| \leq 60 \cdot
10^{-10}$. This might represent the first modern indication for the
existence of a preferred reference frame and for a non-zero
anisotropy of the speed of light. At the same time, since this range
we have found for the RMS parameter is consistent with the
theoretical prediction $\sim 42\cdot 10^{-10}$ of
Refs.\cite{pagano,pla}, we emphasize once more the importance of a
fully model-independent analysis of the data. \vfill\eject

\end{document}